%
%
\input amstex
\documentstyle{fic}
\NoBlackBoxes

\topmatter
\title Toward a Definition of Chaos for General Relativity\endtitle
\author Donald Witt and Kristin Schleich\\
\address Department of Physics\\ University of British Columbia\\
Vancouver, BC V6T 1Z1, Canada\endaddress
\endauthor

\leftheadtext{Donald Witt and Kristin Schleich}
\rightheadtext{Toward a Definition of Chaos for General Relativity}

\cvol{00}
\cvolyear{0000}
\cyear{0000}

\subjclass Primary 70K50; Secondary 83C05\endsubjclass
\abstract 
General relativity exhibits a unique feature not represented in
standard examples of chaotic systems; it is a spacetime diffeomorphism 
invariant theory. Thus many characterizations of
chaos do not work.  It is therefore necessary to develop a definition of chaos 
suitable for application to general
relativity.  This presentation will present results
towards this goal. 
\endabstract

\thanks The second author is supported in part by an NSERC grant.
\endthanks
\endtopmatter

\document

\head 1\enspace Introduction \endhead 

Recent work in chaos has lead to quantitative
studies of chaotic behavior in different mathematical and physical systems. Such 
results have led to useful insight into the nature of these systems. Clearly,
a characterization of whether or not general relativity exhibits chaos would be 
valuable.  However, there are well known difficulties in applying the current 
characterizations of chaos to general relativity; they  are
not diffeomorphism invariant. Thus when applied to general relativity,
they give different results in different coordinate systems (See for example, 
summary and references in \cite{Rugh [1994]}). 
In particular, one test for chaos commonly used by physicists
is to search for sensitive dependence to 
initial conditions in a system $\dot x(t)=F(x)$ by computing Liapunov exponents 
via  $ \lambda =\lim_{t\to\infty}{1\over t} \ln |z(t)|$ where $z(t)$ is the 
linearized perturbation of
$x(t)$. Positive Liapunov exponent is taken as indicating chaotic behavior. 
Application of this test to general relativity is fundamentally flawed; a 
coordinate change rescales a positive Liapunov exponent to zero. For example,
consider  computing the Liapunov exponents for the one parameter
model corresponding to expanding de Sitter spacetime; one finds $\lambda = H$ 
in the  coordinate chart $ds^2= - dt^2 + a^2(t)(dx^2 + dy^2 +dz^2)$  
whereas $\lambda = 0$ in the chart
$ds^2= - {{dt^2}\over{H^2 t^2}} + a^2(t)(dx^2 + dy^2 +dz^2)$.
Clearly, as the physics of the solution is independent of coordinate choice,
there  is a problem with this characterization.

It is clear that a coordinate independent definition of chaos is needed for the 
study of general relativity. This paper discuss work toward developing such a 
definition. A longer version of this paper with details is in preparation 
(\cite{Schleich and Witt [1995]}). The starting point will be a discussion of a 
rigorous definition of chaos for standard dynamical systems. Next crucial 
aspects of general relativity that distinguish it from these standard systems 
will be outlined.  Finally  a generalized definition of chaos
that addresses these crucial aspects will be proposed.

\head 2\enspace Chaos in Standard Dynamical Systems\endhead

The literature contains many characterization of chaos, often adapted to
the type of the system being studied.  However,  
it is well known that testing for chaos simply by
testing for positive Liapunov exponent is flawed. For example,
the system with evolution ${\dot x} =cx$ where $c$ is a  positive constant
has positive Liapunov
exponent. Clearly, this system is not chaotic.
Rigorous
characterizations of chaos require ergodicity in addition to tests of sensitive 
dependence on initial conditions such as Liapunov exponents. 
(See for example, \cite {Wiggins [1990], p. 608} and \cite{Pollicott [1993]} 
in the context of non-uniformly hyperbolic diffeomorphisms.)

While developing the machinery for a rigorous definition of chaos, it is useful
to keep in mind two standard examples of dynamical systems:
One corresponds to two linear coupled oscillators;
for fixed initial momenta, this system is
isomorphic to the translational flow on the flat torus 
$T^2={\Bbb R}^2/{\Bbb Z}^2$, $f_t:T^2\to T^2$ given by
$f_t(x_1,x_2) =(x_1+\omega_1 t,x_2+\omega_2 t)+{K}$, ${ K }\in {\Bbb Z}^2$.
The other is the cat map on the flat torus $T^2$ given
 by $f(x_1,x_2)=(2x_1+x_2,x_1+x_2) +{ K}$, ${ K }\in {\Bbb Z}^2$.

A general dynamical system is a measurable map:
A measure space $(X,{\Cal B})$ consists of  a set $X$ and a $\sigma$-algebra 
$\Cal B$. A map $T:(X_1,{\Cal B}_1) \to (X_2,{\Cal B}_2)$ between
two measure spaces is called measurable if and only if 
$T^{-1}(B) \in {\Cal B}_1$ for all $B\in{\Cal B}_2$.
A measure $m$ on $(X,{\Cal B})$ is a function 
$m:{\Cal B}\to {\Bbb R}^+\cup\{\infty\}$ which satisfies $m(\emptyset)=0$ and
$ m(\cup_k B_k)=\sum_k m(B_k)$ for any countable disjoint collection $\{B_k\}$.
Then
\definition{Definition 2.1} A measure $m$ on $(X,{\Cal B})$ is a probability 
measure if $m(X)=1$.
\enddefinition
\noindent Note that $X$ need not be a compact space; for example the measure 
induced by ${1\over\sqrt{\pi}}\exp(-x^2) dx$ is a probability measure on 
$\Bbb R$. For simplicity of notation, the term measure will refer to
 probability measure in this paper. Next
\definition{Definition 2.2} Given a measurable map $T$ on $(X,{\Cal B})$, $m$
is T-invariant if and only if $T^*m(B)=m(B)$ for all 
$B\in {\Cal B}$ where $T^*m(B) = m(T^{-1}(B))$.
\enddefinition
\noindent When the map is understood, a T-invariant measure is called an 
invariant measure. In general,
a map $T$ on a given space may have more than one invariant measure. 
Especially note that $m$ need not have support on the entire space; for
example $m=\delta(x)$ is an invariant measure for the map $T:\Bbb R \to \Bbb R$ 
given by $T(x) = ax$.  At this point a key concept can be introduced:
\definition{Definition 2.3} An invariant  measure $m$ is called an ergodic
 measure if whenever $T^{-1}B = B$ for some $B\in {\Cal B}$,
then either $m(B)=0$ or $m(B)=1$.
\enddefinition
\noindent Note that in general, the space of ergodic measures is not equal to 
the space of invariant measures.  One can show that when $T$ is a homeomorphism, 
there exists at least one ergodic measure on any compact space. 
Both examples have ergodic measures;
the measure induced by $dx_1dx_2$ is the unique invariant and ergodic measure 
for the cat map and for the translational flow when
${\omega_1\over\omega_2}  \not\in \Bbb Q$. 
For  ${\omega_1\over\omega_2} \in \Bbb Q$, there are an {\it infinite} number of 
ergodic measures for the translational flow; these
 correspond
to delta functions that concentrate support on one periodic orbit. 
 A  rigorous definition of chaos is then
\definition{Definition 2.4} A system $T:(X,{\Cal B})\to (X,{\Cal B})$ is chaotic 
if and only if  it has an ergodic measure $m$
and exhibits sensitive dependence on initial conditions with respect to
this measure.\footnote{One test for sensitive dependence
is now given by positive Liapunov exponents {\it subject} to computation in the 
ergodic measure $m$ (See for example Pollicott [1993] for application to 
nonuniformly hyperbolic diffeomorphisms). This definition differs significantly 
from the standard physicist's use in that Liapunov exponents  defined w.r.t. 
an ergodic measure are constant over the space.  Alternately 
\cite{Wiggins [1990]} presents a definition that does not presume 
exponential divergence.}
\enddefinition

\noindent 
When applied to the examples, (2.4) distinguishes chaotic behavior.
The translational flow is not chaotic for any values of the parameters, even 
incommensurate ones; it is easy to show that trajectories do not
diverge by any test of sensitive dependence as they are simply straight lines. 
In contrast, the cat map is chaotic as it has positive Liapunov exponent; 
it illustrates the importance of sensitive dependence in distinguishing chaotic 
behavior. 

Note especially that the use of ergodic measure implies that chaotic behavior can 
be concentrated on a subset of the measure space. This feature is important as 
dissipative systems such as the Henon map exhibit chaotic behavior on such a 
subset. Thus the use of the ergodic measure make (2.4) a very general definition 
of chaos. Moreover, other characterizations of chaos
such as self-similarity and dense sets of periodic orbits can be recovered under 
certain conditions. It is this definition that will be the starting point for a 
more general definition of chaos suitable both to the standard examples
and to general relativity.

\head 3\enspace Important Characteristics of General Relativity\endhead

There are two issues that must be addressed by any formulation of chaos 
appropriate for application to general relativity: its diffeomorphism 
invariance and its lack of a physically motivated choice of measure.

The most discussed difficulty associated with diffeomorphism invariance is 
concentrated in time reparameterization invariance; time is no longer an 
absolute quantity but is tied to the coordinate system.  The coordinate systems 
given for the de Sitter space model clearly indicate this fact. Note that one can 
always associate an observer (though not necessarily a freely falling observer) 
to any choice of coordinate system; thus any attempt to tie chaos to observers 
is simply making physical a particular coordinate choice.

This problem is closely linked to the fact that
diffeomorphism invariance is a form of gauge invariance; the metric
degrees of freedom are not all physical. However, it is key to the correct 
identification of chaos that it be carried out on physical degrees of freedom 
alone. For example, consider the lifts of the translational flow and the cat map 
from $T^2$ to the covering space ${\Bbb R}^2$.
 The measure induced by $dx_1dx_2$ is not ergodic
for either map on ${\Bbb R}^2$; there are obviously invariant sets for both the 
translational flow  and the cat map. Therefore neither system is ergodic on the 
covering space in this measure. 
Clearly the global topology of the space plays a key role in the nature of
the dynamical system! 

Such global topology also arises in general relativity; 
removing the large gauge transformations will change the
global structure of the phase space. An example is provided by Bianchi I  
(\cite{Bhushan et. al. [1995]}). Spatial gauge is
fixed locally by writing the metric as 
$ds^2 = -N^2(t)dt^2+a^2(t)dx^2 + b^2(t) dy^2 + c^2(t) dz^2$. 
Residual transformations corresponding to interchanges of spatial coordinates
are not fixed. Removing these modifies the nature of the flow; 
the fully reduced phase space is no longer ${\Bbb R}^2$ but a wedge and
the flows are no longer straight lines but include bounces
at the boundaries of this wedge. 
This change in behavior in a such a simple model stresses the need to fully 
reduce to physical degrees of freedom.

Unfortunately, this problem is not confined to large gauge transformations; note
that the usual
Hamiltonian formulation of  gravity in terms of superspace is also {\it not} in 
terms of physical degrees of freedom. Although the spatial diffeomorphisms have
been removed, those corresponding to timelike diffeomorphisms have not. 
Therefore the study
of flows on this space includes redundant information unless a further reduction
is carried out. In particular as different time reparameterizations correspond 
to different gauge
choices, studying the dynamics of the system without further reduction imparts 
physical meaning to different gauges. Thus gauge effects can mask the underlying 
physics of the system. Unfortunately there are no known ways to reduce to 
physical degrees of freedom explicitly; one obvious approach is to isolate a 
preferred timelike Killing vector or conformal Killing vector for superspace. 
However, \cite{Kuch\v ar [1981]} has shown
no such vectors exist in general. It is thus unlikely that explicit reductions
can be carried out. 

The second issue is the lack of a physically motivated choice of measure
for general relativity. It is well known that there is no preferred metric
on superspace. Therefore, one cannot begin with the corresponding measure
 as a candidate for an invariant or ergodic measure.
 This problem is also complicated by the fact that
superspace is not in terms of physical degrees of freedom. Thus it is not clear 
what role any choice of metric  would play in defining any such measure.

\head 4\enspace Toward a Generalized Definition of Chaos\endhead

The above issues clearly demonstrate the futility
of testing for chaos in general relativity by (2.4) by studying flows on 
superspace. However, what should
this definition be replaced by? First observe that general relativity is a 
hamiltonian system in terms of its physical degrees of freedom, even if these 
variables cannot be identified explicitly.
It is therefore natural to first ask the question in the
context of the fully reduced theory, even if it cannot be explicitly constructed.
Given such a definition on the physical
degrees of freedom, one can then attempt to parameterize it to be applicable to 
more tractable forms of the theory.

Observe that a key element in (2.4) is ergodicity,
intuitively that  orbits in the phase space are dense in some measurable set.  
Topological generalizations of ergodicity are well known (See for example 
\cite{Peterson [1990]}). Therefore, a natural approach is to generalize the 
condition of ergodic measure to a topological condition.

Given any invariant measure $m$ and a map $T$ one can define measurable entropy
$0\le h_{meas}(T,m) $ (See \cite{Peterson [1990]} sect. 5.1).
The measurable entropy is related to the (rigorous) Liapunov exponents by the 
Pesin-Reulle inequality under certain restrictive
conditions on the map and measure
$h_{meas}\le\sum_{{\lambda_i>0}} \lambda_i$
where the sum is over positive Liapunov exponents 
(\cite{Pollicott [1993] p. 46}).
Therefore $h_{meas}>0$ implies chaos for ergodic measures. Next
 \definition{ Definition 4.1} The topological entropy is
$h_{top}(T) = \sup_m h_{meas}(T,m)$ where $m$ is an invariant measure.
\enddefinition
\noindent Clearly if there is any measure with $h_{meas}>0$, then $h_{top}>0$. 
This is independent of whether or not the measure has support 
on the entire space.
The utility of this definition of entropy is given in the following theorem
\proclaim{Theorem 4.2} If the topological entropy of a map $T$ is positive, 
then there exists an ergodic
measure such that the measurable entropy is positive. 
\endproclaim
\noindent Thus topological entropy encompasses both ergodicity and sensitive 
dependence in one quantity.
The translational flow has zero topological entropy
by explicit calculation. The cat map has positive topological entropy; it is 
a special case of a family of chaotic systems 
 with positive topological entropy given by maps
 on a 2-manifold of $\hbox{\rm genus} > 1$,
$T_*:H_1(M^2;{\Bbb R}) \to H_1(M^2;{\Bbb R})$
such that $T_*$ has all $\hbox{eigenvalues} >1$.
The difficulty with topological entropy has  been in its
application. However, numerical estimates of this quantity 
are possible (\cite{Gribble [1995]}).

Given the above, a natural generalized definition of chaos is
 \definition{ Definition 4.3} A dynamical system is chaotic if and only 
 $h_{top}>0$.
\enddefinition
\noindent This definition is especially suited to systems without known 
ergodic measure such as general relativity.

\head 4\enspace Conclusions\endhead

Definition (4.3) can be taken as a working definition of chaos for general
relativity; it is not a final definition but a plausible starting point. Its
difficulty is that it must be applied to the theory in terms of physical degrees 
of freedom. A key task is to see if it can be recast into parameterized form. 
Such a hope is not unreasonable given that
topological entropy is closely related to partition functions which have obvious
analogs in gauge theories.  Furthermore
quantization of gauge theories in general proceeds along a very similar track of
reparameterization of quantities formally given in terms of physical degrees of 
freedom.
Finally, infinite dimensional systems such as general relativity
only have finite evolutions, that is singularities
and caustics can arise in the future evolution of regular initial data.
 These effects do not appear in finite dimensional systems. Although such effects 
 appear
in other infinite dimensional theories, they are an essential part of the nature
of general relativity. Thus it must be verified that (4.3) correctly handles 
these features.

\refstyle{B}

\enddocument